# TANGENTIAL MODEL OF MICROTUBULES

S. Zdravković and S. Zeković

*Univerzitet u Beogradu, Institut za nuklearne nauke Vinča, 11001 Beograd, Serbia. (szdjidji@vin.bg.ac.rs)*

**ABSTRACT**
Microtubules (MTs) represent basic components of a cytoskeleton. The present work studies nonlinear dynamics of MTs assuming tangential oscillations of the dimers. We introduce a two component model and show that the dynamics of MTs can be explained in terms of breather solitary waves.

**INTRODUCTION**
MTs are the basic components of the cytoskeleton, existing in eukaryotes [1]. They are long structures that spread between a nucleus and cell membrane, playing essential role in the shaping and maintenance of cells and in cell division. Also, they represent a network for motor proteins. More information about MT structure and function can be found in a review [2].

   MT is a hollow cylindrical polymer. Its surface is formed usually of 13 long structures called protofilaments (PFs), representing a series of electric dipoles called dimers. For most of the models the dimer`s internal structure is not taken into consideration.

   There are a few models describing the nonlinear dynamics of MTs, based on the crucial fact that MTs are ferroelectrics as the dimers are electric dipoles [3]. Each model can be seen as either longitudinal or angular. The first nonlinear model was longitudinal one, introduced almost 30 years ago [3]. Its improved and more general version is the so-called $u$-model [2,4]. Two commonly used mathematical procedures were explained within the $u$-model in Ref. [5].

**TANGENTIAL MODEL OF MICROTUBULES**
An overview of a couple of models, both longitudinal and angular, can be found in the review [5]. All of them are one component models. The purpose of this paper is to introduce a two component angular one. It was pointed out that the MT represents a ferroelectric system. Interaction of a single dimer with surrounding ones, that do not belong to the same PF, can be modelled by W-potential energy, or potential for short [3]. This is a function $f_W = -ax^2 + bx^4 - cx$, having two minima. We assume that all parameters are positive. It is crucial to understand the meaning of these minima. The existence of the two minima means that there are two possible positions for the dimer. To be more precise, there are two directions of electric fields, $\vec{E}_1$ and $\vec{E}_2$, around which the dimer can oscillate. One of them is shown in Fig. 1. A resultant internal electric field $\vec{E} = \vec{E}_1 + \vec{E}_2$ is in the direction of PF. We may expect that the dimer oscillates around $\vec{E}$, but any displacement would move it towards the directions of either $\vec{E}_1$ or $\vec{E}_2$. This means that the dimer's position in the direction of $\vec{E}$ is not stable and corresponds to the maximum of the function $f_W$.

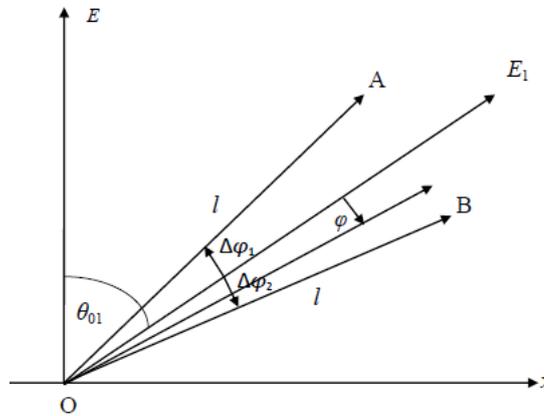

The values of the longitudinal $\vec{p}_z$, radial $\vec{p}_r$, and tangential $\vec{p}_\theta$ components of electric dipole moment are known [6]. The component $p_z$ is in the direction of MT. The oscillation of the dimer is in the tangential plane, which means that $\vec{p}_r \cdot \vec{E}_1 = 0$ and the relevant moment here is $p = \sqrt{p_z^2 + p_\theta^2}$.

**Figure 1.** A schematic representation of the dimer's oscillations.

In Fig. 1, the dimer oscillates around the direction of $\vec{E}_1$. The angle between PF and $\vec{E}_1$ is $\theta_{01} \equiv \theta_0$. A coordinate determining the displacement from the direction of $\vec{E}_1$ is $\varphi$. Its position with respect to the direction of PF is $\theta$. The amplitude positions of the dimer are OA and OB. Hence, $\Delta\varphi_1 \leq \varphi \leq \Delta\varphi_2$, $\Delta\varphi_1 < 0$ and $\Delta\varphi_2 > 0$. The value $\theta_0$ corresponds to the minimum of the W-potential [7].

The Hamiltonian for MT can be written as [7]

$$H = \sum_n \left[ \frac{I}{2}\dot{\varphi}_n^2 + \frac{k}{2}(\varphi_{n+1} - \varphi_n)^2 - \frac{A}{2}\theta_n^2 + \frac{B}{4}\theta_n^4 - C\theta_n - pE_1 \cos\varphi_n \right]. \quad (1)$$

We recognize the kinetic energy, potential energy of the interaction between neighbouring dimers belonging to the same PF in the nearest neighbour approximation, and W-potential. The potential energy which would explain the interaction between neighbouring dimers belonging to different PFs is neglected [8]. The very last term comes from the fact that the dimer is the electric dipole existing in the field of all other dimers, where $p > 0$ is an electric dipole moment. It is assumed that $E_1 > 0$ [7].

From $\theta = \theta_0 + \varphi$ and Eq. (1) we obtain the dynamical equation of motion $I\ddot{\varphi}_n = k(\varphi_{n+1} + \varphi_{n-1} - 2\varphi_n) - A_0\varphi_n - C_0\varphi_n^2 - B_0\varphi_n^3$, where $B_0 = B - pE_1/6$, $A_0 = -A + 3B\theta_0^2 + pE_1$, $C_0 = 3B\theta_0$ [7]. To solve it, we use a semi-discrete approximation [9]. This means that we assume small oscillations, i. e. $\varphi_n = \varepsilon \Phi_n$, $\varepsilon \ll 1$, and look for wave solutions in the form

$$\Phi_n(t) = F_1(\xi)e^{i\theta_n} + \varepsilon \left[F_0(\xi) + F_2(\xi)e^{i2\theta_n}\right] + cc + O(\varepsilon^2), \qquad (2)$$

where $\xi = (\varepsilon nl, \varepsilon t)$, $\theta_n = nql - \omega t$, $\omega$ is the optical frequency of the linear approximation, $q = 2\pi/\lambda$ is a wave number, cc represents complex conjugate terms and $F_0$ is real. A crucial point is that the function $F_1$ represents an envelope, which is treated in a continuum limit, while $e^{i\theta_n}$, including discreteness, is a carrier component of the wave. After rather tedious mathematics, including a continuum limit $nl \to z$, we come up with a conclusion that the functions $F_0(\xi)$ and $F_2(\xi)$ can be expressed through $F_1(\xi)$, while $F_1(\xi)$ is a solution of the solvable nonlinear Schrödinger equation (NLSE) [9]. The final result, that is the solution of the equation of motion, is (work in progress):

$$\varphi_n(t) = 2A'\operatorname{sech}(\Phi_1)\left\{\cos(\Phi_2) + A'\operatorname{sech}(\Phi_1)\left[\mu/2 + \delta\cos(2\Phi_2)\right]\right\}, \qquad (3)$$

where $\Phi_1 = (nl - V_e t)/L$ and $\Phi_2 = \Theta nl - \Omega t$. A very tedious parameter selection can be found in Ref. [7]. It suffices now to mention that the wave velocity $V_e$ can be obtained using the idea of a coherent mode (CM), assuming that the envelope and carrier wave velocities are equal, i. e. $V_e = \Omega/\Theta$. This means that the function $\varphi_n(t)$, shown in Fig. 2 for CM and a certain set of the parameters [7], is the same at any position $n$. This is a modulated solitary wave called a breather.

**CONCLUSION AND FUTURE RESEARCH**

The model explained here is a mechanical one. An example that shows the relevance of the mechanical models is kinocilium, a component of vestibular hair cells of the inner ear [10]. Also, MTs are believed to be a source of electrodynamics activity of cells and have been modelled as nonlinear RLC transmission lines, which is very important for fighting some diseases [11]. Therefore, MTs are both mechanical and electrical systems. Regarding their modelling, the best that should be done is to work towards more component models that would take both characteristics into consideration [12]. Our approach here is a classical one and NLSE should not be confused with the quantum SE.

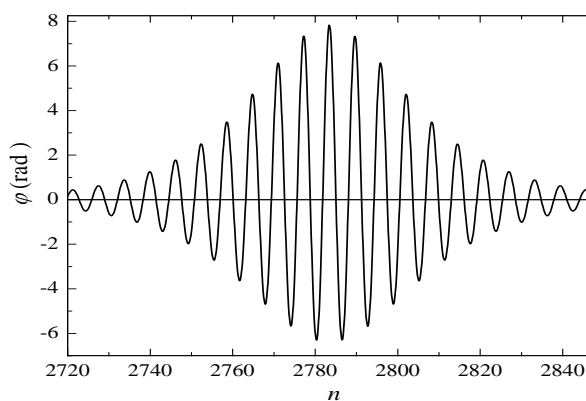

**Figure 2.** The function $\varphi_n(t)$ as a function of $n$ for $t = 50\,\text{ns}$. The CM is assumed.

*Acknowledgement*

We thank our colleague Alexandr N. Bugay for his collaboration with us.

This article is based on a contribution given at the conference:

**PHYSICAL CHEMISTRY 2021**
15th International Conference on Fundamental and Applied Aspects of Physical Chemistry
Proceedings
Volume I
September 20-24, 2021
Belgrade, Serbia